\documentstyle[12pt,epsfig]{article}
\begin{document}
\begin{titlepage}
\vspace*{-1cm}
\begin{flushright}
CERN--TH/97--182
\end{flushright}                                

\vskip 1.cm
\begin{center}          
{\Large\bf MINIJETS AS A PERTURBATIVE PROBE OF COLOUR CHARGE}
\vskip 1.3cm
{\large David J. Summers }
\vskip .2cm
{TH Division, CERN, CH-1211 Gen\`eve 23, Suisse}
\vskip 6cm   
\end{center}   
\begin{abstract}
We motivate the study of ``Minijets'' (that is jets soft with respect
to the hard scattering, but hard with respect to $\Lambda_{QCD}$) as a
means to study the underlying QCD colour flow in events. We discuss,
with the aid of a simplistic model,
currently available events in which minijets are observed, that is the
events have high multiplicity, as a means of understanding the physics
of minijets. We encourage both theoreticians and experiments to
continue the study of minijets.
\end{abstract}                                                                
\vskip 5cm
Contribution to 32nd Rencontres de Moriond: QCD
and Hadronic Interactions
\vfill
\begin{flushleft}
CERN--TH/97--182
\end{flushleft}
\end{titlepage}
\newpage

In the past the concept of QCD colour flow has been an interesting
prediction, and hence test, of QCD. One of the more surprising
examples of QCD colour flow has been the rapidity gaps seen
at HERA \cite{rapgap}. In the majority of jet events observed at HERA
the incoming 
electron/positron radiates a photon, which scatters off one of the
coloured constituents of the proton, to produce the observed jet.
This hard scattering leaves the proton remnant and the pre-jet
in a non colour singlet state, during hadronization the colour
of these objects is rearranged to form colourless hadrons, and these
hadrons fill the gap between the produced jet and the proton remnant
with an underlying event.
However among the HERA events are some with an observed jet produced
at central rapidities; but with no hadrons between the jet and proton
remnant direction; {\em i.e.} there is no underlying event, and this means
that the mechanism for producing this jet can not be that just
described. The lack of hadrons between the observed jet and the proton
remnant means that there is no QCD colour is exchanged
between the proton and the photon, and hence that photon must have
scattered off a colour 
singlet object in the proton, the pomeron. So the flow of hadrons
in an event has told us something about the structure of the proton,
that the struck object inside the proton that took part in the hard
scattering is a QCD colour singlet.

\begin{figure}
\begin{center}
\psfig{figure=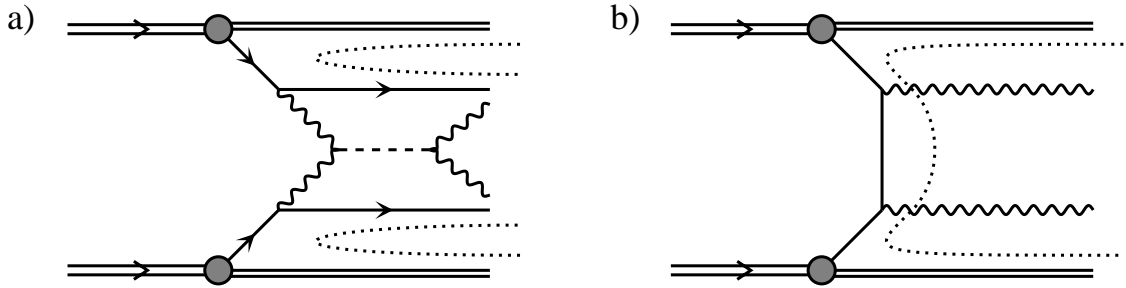,height=1.5in}
\end{center}
\caption{Typical Feynman diagrams for $pp$ scattering to a pair of vector
  bosons. The a) signal, vector boson scattering via a
  Higgs boson, b) background, divector boson production. 
  Show, dotted, is the leading QCD colour flow. 
\label{fig:wwh}}
\end{figure}

This ability to probe the QCD colour of the object that takes part in
the hard scattering is, in the future, likely to provide a strong
experimental probe. A prime example is in the detection of a heavy
($\sim 500$ GeV) Higgs boson at the LHC, such a Higgs is produced via
two different mechanisms. Approximately two thirds of heavy Higgs are
produced from $gg$ fusion via a top quark loop, with the remaining
third produced via vector boson fusion, show in Fig.\ref{fig:wwh}a. 
Such a heavy Higgs typically
decays into two massive vector bosons, either $W$ or $Z$ bosons, and
the back ground from continuum diboson production, shown in
Fig.\ref{fig:wwh}b, can be large. However the QCD colour flow in the
diboson fusion signal process is very different from the background
continuum diboson production, the leading colour flow is shown in
Fig.\ref{fig:wwh} as a dotted line. For the diboson fusion process
QCD colour flows between the proton remnant and the parton that emits
the vector boson, this parton travels in very much the same direction
as the proton remnant. This means that the hadrons associated with the
underlying event will travel in a similar direction to the proton
remnant, i.e. at very large rapidities. For the diboson continuum
background (as well as the Higgs signal from $gg$ fusion) QCD colour
is exchanged between the two protons, and this means that the
underlying event will produce hadrons at all rapidities. 
This difference in the flow of hadrons gives us a method of separating
the vector boson fusion Higgs signal from the continuum background,
effectively the Higgs signal has a double rapidity gap on both sides of
the event, with the Higgs decay taking place at central rapidities
\cite{wwh}.

However this method for separating signal from background by the flow
of hadrons has many difficulties. On the theoretical side 
we make predictions and the rate of the signal and background based
upon perturbative QCD (pQCD). pQCD's fundamental fields are partons
(quarks and gluons) and it is about these fields that pQCD makes
predictions. By local parton hadron duality we expect that the
produced partons will hadronize into jets; however this tells us
nothing about individual hadrons and it is in terms of individual
hadrons that rapidity gaps are defined. This means that we can make no
absolute predictions based upon pQCD of the signal rate of events
where a rapidity gap is observed, or indeed the background rate where 
a fluctuation leads to an observed rapidity gap, despite QCD colour
being exchanged between the protons. On the experimental side the
cross section for Higgs productions drops very rapidly as the Higgs
gets heavy, and for very heavy Higgs the cross section is small enough
that it is only with very high luminosities that we produce any Higgs
events at all. However at very high luminosities there are likely to
be many interactions per bunch crossing, and each of these
interactions will produce its own underlying event, which will wash
out the rapidity gap at central rapidities.

One possible solution to these problems is rather that look for events
with a rapidity gap, {\em i.e.} no hadrons at central rapidities, instead to
look for events with no soft ``minijets'' at central rapidities
\cite{pertwwh}.
Naively we expect additional jets to be suppressed by a factor of
$\alpha_s$, however there can often be a large volume of phase space
in which additional jets can be radiated, and this can enhance the
additional jet rate by a large logarithm such that multijet emission
is not suppressed. In particular jet emission that is soft with
respect to the hard scattering is enhanced by
such a large logarithm. If we define soft jets in such a way that the
background divector boson continuum (which has more QCD colour
acceleration, and hence more jet 
activity) typically has additional jet emission, while the vector boson
fusion signal (with less QCD colour accelerated) typically has no
additional jet activity; then the difficulties with using hadrons to
define a rapidity gap are overcome. Additional soft jet production we
expect to be predictable by pQCD as long as the scale for soft jet
emission is far above $\Lambda_{QCD}$. On the experimental side we
expect that the majority of interactions will be very soft, and
although producing an underlying event the majority of interactions
will not produce observed jets, and so not interfere with tagging a
Higgs signal event. This means that firm theoretical predictions can,
in principle, 
be made about a minijet tag of a Higgs signal, and that a minijet tag
can be used at high luminosities where there are overlapping events. 

\begin{figure}
\begin{center}
\psfig{figure=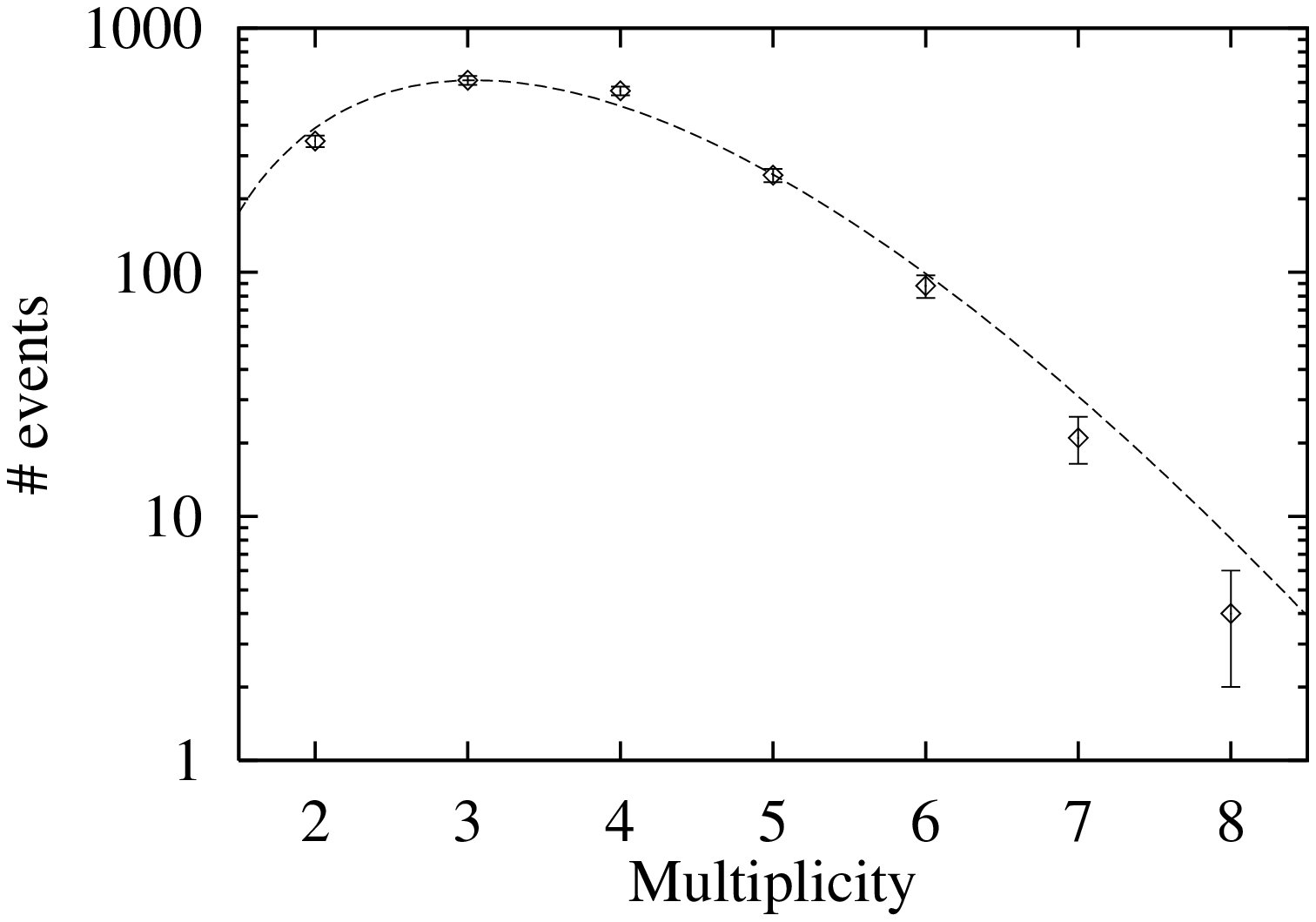,height=3.in}
\end{center}
\caption{The multiplicity distribution for CDF events with cuts show
  in equation \ref{eq:cuts}. The errors shown are purely
  statistical. The plotted curve is proportional to 
  $P_n(\bar n) = {\bar n^n\over n!}\; e^{-\bar n}$ where $n$ is the
  number of jets observed over the two that are forced by kinematics
  and $\bar n$ is the average additional muliplicity, $\bar n = 1.57$. 
\label{fig:mult}}
\end{figure}

However before we can use minijet emission as a probe of QCD colour
charge it is necessary to understand minijet emission both on a
theoretical and experimental level. In order for multiple jet emission
to be important we need a large phase space for emission of those
jets, and this means that we require the hard scattering to be as
energetic as possible. Currently the most energetic events are to be
found at the TeVatron, and CDF have studied the multijet distributions
in there most energetic events \cite{CDFmultijet}. They apply the cuts,
\begin{eqnarray}
E_{T \rm jet} > 20~{\rm GeV} &\qquad , \qquad& 
                |\eta|< 4.2 \nonumber\\
\sum E_T > 420~{\rm GeV} &\qquad , \qquad& 
                m_{\rm jets} > 600~{\rm GeV} \nonumber\\
|\cos\theta^*|< {2\over 3} &\qquad  \qquad&
\label{eq:cuts}
\end{eqnarray}
where $\theta^*$ is the scattering angle of the 
highest $E_T$ jet in the multijet centre of mass frame.
Due to the high invariant mass of these events the events are forced
to be fairly central, and this means that at least two jets are always
observed. In figure \ref{fig:mult} we show the multiplicity
distribution observed in the CDF events, although 3rd and additional
jets are naively produced at a higher order in perturbation theory
because of the large volume of phase space such additional jet
activity is not suppressed, with typical events with 3 or 4 jets
observed.

The high multiplicity of these events means that pQCD in its lowest
order is not not reliable. The large logs that give rise to the high
multiplicity need to be dealt with in the theoretical calculation.
Now most of the additional jet activity takes place at low $E_T$, and
these soft jets typically originate from gluons. Now for soft gluons
the matrix element factorises, the probability to radiate $n$
additional soft gluons is given by,
\begin{equation}
|{\cal M}_{ng}^2| = |{\cal M}_1|^n \; .
\end{equation}
The phase space to radiate $n$ additional gluons is given by,
\begin{equation}
d({\rm LIPS}_n) = {1 \over n!} \prod^n {d^3\over (2\pi)^3 2 E}
                    \delta^4(P-\sum p) 
                  H(\hbox{observed properties of the gluons})
\end{equation}
where the $1/n!$ term arises from the symmetry factor for integrating
$n$ gluons over the same phase space. Now if $H = 1$, that is we make
no experimental requirements about the observed properties of the
gluons, then in moment space we find,
\begin{equation}
d({\rm LIPS}_n) = {1 / n!} \prod^n d({\rm LIPS}_1) \; .
\end{equation}
This means that the probability to see $n$ gluon jets is given
by,
\begin{equation}
{\cal P}(\hbox{n gluons}) = {\cal P}(\hbox{1 g})^n/n! \label{eq:atz}
\end{equation}
and so,
\begin{equation}
{\cal P}(\hbox{any number of gluons}) = \exp({\cal P}(\hbox{1 g})) \;.
\end{equation}
However for the case of jets defined by the CDF algorithm $H\neq 1$,
also not all jets are soft, and not all jets are gluon initiated. This
means that this simple factorisation of the additional jet rate does
not hold, however it is a reasonable ansatz to assume equation
\ref{eq:atz}. 

This is not the full story though, equation \ref{eq:atz}
does not conserve probability, this happens because the rate to
radiate an addition gluon ${\cal M}_1$ is an inclusive quantity, whereas
if we are to make predictions about the specific multiplicity
distributions then we require the exclusive rate to produce 
{\em exactly} $n$ additional jets. We can obtain the exclusive jet
rate by adding a Sudakov suppression factor $\exp(-{\cal P}(\hbox{1 g}))$
\cite{sudakov} that excludes all additional jet activity, and then
equation 
\ref{eq:atz} produces exactly $n$ exclusive jets. This gives the
ansatz for additional jet activity over the hard scattering as,
\begin{equation}
{\cal P}(\hbox{exactly n additional jets}) = 
   {\cal P}(\hbox{1 additional jet})^n/n!\; 
   \exp(-{\cal P}(\hbox{1 additional jet})) \label{eq:satz}
\end{equation}
where ${\cal P}(\hbox{1 jet})$ is the average multiplicity $\bar n$,
and at the leading logarithmic level is given by,
\begin{equation}
{\cal P}(\hbox{1 jet}) = {\sigma(\hbox{3 jet production})^{\rm LO}
                          \over \sigma(\hbox{total 2 jet production})}\;.
\label{eq:pfit}
\end{equation}
This formula has the advantage over strictly tree level pQCD of being
still valid even when ${\cal P}(\hbox{1 jet}) > 1$. Also we have,
\begin{equation}
\sigma_3=\sigma_2 {\cal P} \exp(-{\cal P} )
  = \sigma_3^{\rm LO}(1+{\cal O}(\alpha_s))
\end{equation}
and so the 3 jet cross section has the same accuracy as at leading
order.

If we compare equation \ref{eq:satz} with the measured CDF multiplicity
distribution, fitting only ${\cal P}$ to the observed average excess
multiplicity over the 2 hard jets always observed we find the curve
shown in figure \ref{fig:mult}. Although the point for multiplicity
4 is 3.5 standard deviations above the fitted curve, and thus the
equation \ref{eq:satz} has limitations, it is clear that the fit has
got the gist of the multiplicity distribution correct and gives a
place to start when studying event multiplicities, and hence the
physics of minijets.

\begin{figure}
\begin{center}
\psfig{figure=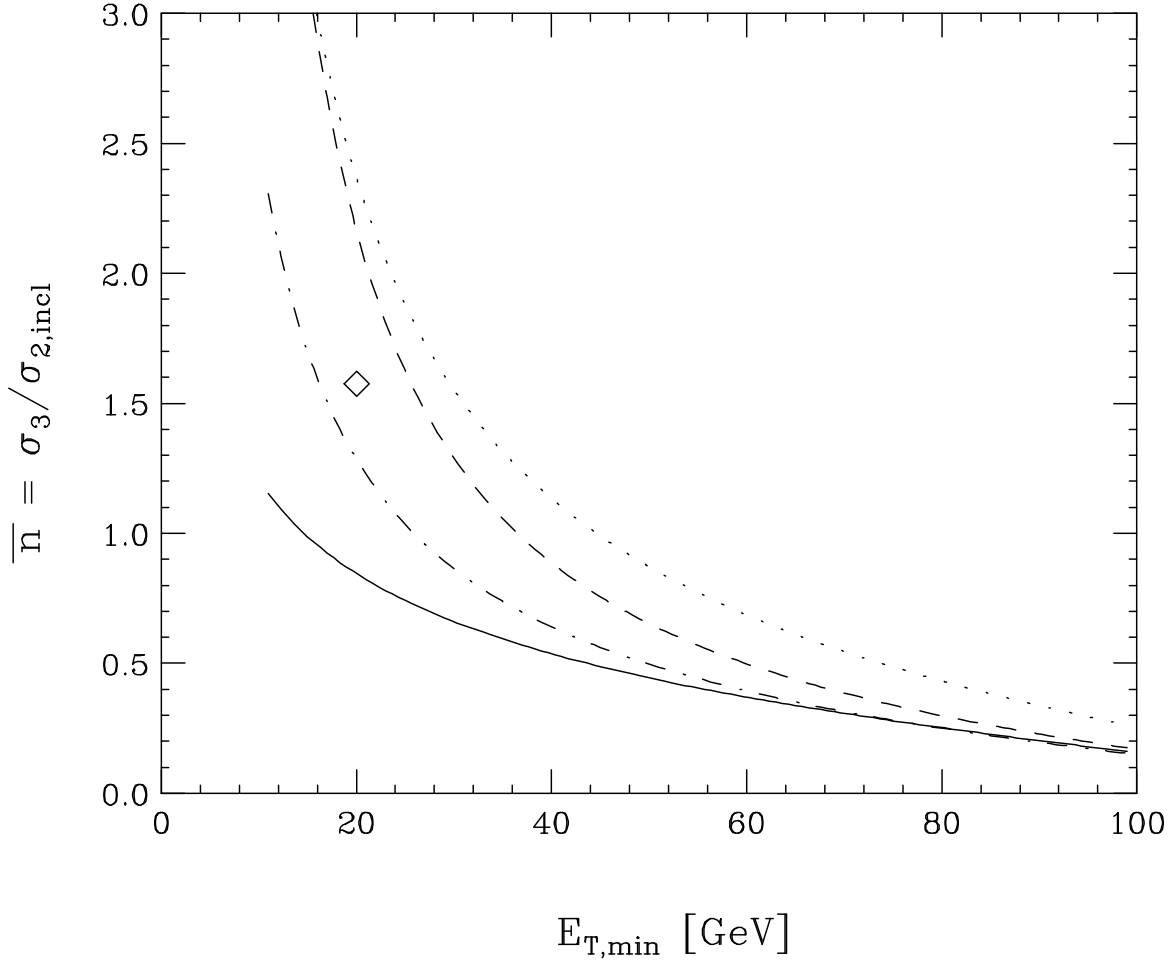,height=5.in}
\end{center}
\caption{Ratio of the tree level 3-jet cross section to the NLO
  cross section for 2-jet inclusive events within the CDF acceptance
  cuts, equation \ref{eq:cuts}. The cross section 
  ratio $\bar{n} = \sigma_3(E_{T,{\rm min}})/\sigma_{2,{\rm incl}}$,
  with $\sigma_{2,{\rm incl}}=33$~pb, is shown as a function of the 
  transverse energy threshold, $E_{T,{\rm min}}$, of the third jet.
  Results are given for four different 
  scale choices in $\sigma_3$: $\mu_R=\mu_F=\sum E_T/4$ (solid line),
  $\mu_R=\mu_F= E_{T,3}$ (dashed line), and $\mu_F=\xi E_{T,3}$, 
  $\alpha_s^3 = \prod_{i=1}^3 \alpha_s(\xi E_{T,i})$ with a scale factor
  $\xi=1$ (dash-dotted line) and $\xi=1/2$ (dotted line). The CDF value for
  the average minijet multiplicity, $\bar{n}=1.57$, is given by the diamond.
  \label{fig:nbar}}
\end{figure}

If we now try to predict the value for ${\cal P}$ from pQCD using
equation \ref{eq:pfit} with a tree level calculation for $\sigma_3$
and a next-to-leading-order calculation for $\sigma_2$ \cite{ggk} then
we find the results shown in figure \ref{fig:nbar} plotted against the
minimum $E_T$ used to define a jet \cite{rsz}.

For the leading order calculation of $\sigma_3$ we have several
different choices of scale. For the renormalisation scale of
$\alpha_s$; this can either be some multiple of the hard scattering
scale, $\sum E_T$,  or some multiple of the soft scale at which the
jet is emitted $E_{T,3}$. For the factorisation scale at which the
parton densities are evaluated we similarly can choose a hard scale
related to $\sum E_T$, if we feel we should sum up the initial state
radiation, or a soft scale related to $E_{T,3}$ if we feel that
explicitly calculating high order emission in $\sigma_3$ means that we
should only generate higher order corrections bellow that
scale. A case can be made for each of these choices, and from
figure \ref{fig:nbar} we can see the prediction for $\bar n$ changes
by a factor of 3 for the experimental value of $E_{T,\rm min}$. This
happens because we have two natural scales in the calculation of
$\sigma_3$ the hard scattering scale, and the $E_T$ of the emisson,
and as these have very different values the predictions for $\bar n$
vary accordingly. The theoretical predictions for $\bar n$
span the experimental measurement, but we can not make an accurate
theoretical predition at this time.

Without improvements to the theoretical predictions we do not
currently have the ability to accurately predict the rate of minijet
emission, and this means that we do not currently have the technology
to use minijets as a absolute theoretical prediction to probe
QCD colour charge. On the theoretical side the large scale dependence
of $\sigma_3$ can be cured by working at higher order in perturbation
theory \cite{fzs}; however this would necessitate an improvement in the
simple minijet model, equation \ref{eq:satz}, which is non consistent
we such a higher order calculation. Alternatively it may be possible
to show which logarithmic terms need to be resummed in order to remove
the large scale dependence, without a full calculation of the higher
order finite corrections. In addition equation \ref{eq:satz} does not
take advantage of our knowledge of the tree level rate for high jet
multiplicities \cite{njets}. Clearly there is much theoretical work that
remains to be done. There is also much work to be done on the
experimental side to help our understanding of minijets, for example
the different theoretical scale choices not only affect the average
multiplicity, but also affect the shape of distributions. This can be
seen in figure \ref{fig:nbar} where a renormalisation scale related to
the $E_T$ of the emitted jet has the average multiplicity growing far
more rapidly as the $E_T$ cut used to define a jet is decreased, than a
harder renormalisation scale. If the experimenter were able to measure
this distribution then this information can be used to motivate
different choices for the scale choice.

\section*{Acknowledgments}
I would like to thank Dieter Zeppenfeld and David Rainwater for their
fruitful collaboration on this project.


\begin{thebibliography}{99}

\bibitem{rapgap} ZEUS collaboration, Phys.~Lett.~B {\bf 315} 481
  (1993), \\
  H1 collaboration Nucl.~Phys. {\bf B429} 477 (1994).

\bibitem{wwh} Yu.L. Dokshitzer, V.A. Khoze and S.I. Troyan,
  Sov.~J.~Nucl.~Phys {\bf 46}, 712 (1987), \\
  Yu.L. Dokshitzer, V.A. Khoze and T. Sj\"ostrand, Phys.Lett.B {\bf
  274} 116 (1992).

\bibitem{pertwwh} V. Barger, R.J.N. Phillips, D. Zeppenfeld 
Phys.Lett.B {\bf 346} 106 (1995).

\bibitem{sudakov}
V. Sudakov, Sov.~Phys.~JEPT {\bf 3}, 65 (1956).

\bibitem{CDFmultijet}
CDF Collaboration, F.~Abe et al., Phys.\ Rev.\ Lett. {\bf 75}, 608 (1995).

\bibitem{ggk} W.T. Giele, E.W.N. Glover, and D. A. Kosower,
Nucl.~Phys.~{\bf B403}, 633 (1993). 

\bibitem{rsz} D. Rainwater, D. Summers, and D. Zeppenfeld 
Phys.Rev. {\bf D55} 5681 (1997).


\bibitem{fzs} S. Frixione, Z. Kunszt, and A. Signer 
Nucl.Phys.{\bf B467} 399 (1996).

\bibitem{njets} F.A. Berends, and H. Kuijf, 
Nucl.Phys.{\bf B353} 59 (1991). 

\end{thebibliography}
\end{document}